# Anisotropic manipulation of terahertz spin-waves by spin-orbit torque in a canted antiferromagnet

T. H. Kim[1,3†], Jung-Il Kim[3], Geun-Ju Kim[3], Kwang-Ho Jang[3], and G.-M. Choi[1,2†]

[1]*Department of Energy Science, Sungkyunkwan University, Suwon 16419, South Korea.*
[2]*Center for Integrated Nanostructure Physics Institute for Basic Science Suwon 16419, South Korea*
[3]*Korea Electrotechnology Research Institute, Ansan 15588, South Korea*

[†]Correspondence and requests for materials should be addressed to T. H. Kim (thkim23@keri.re.kr) and G.-M. Choi (gmchoi@skku.edu).

We theoretically and numerically elucidate the electrical control over spin waves in antiferromagnetic materials (AFM) with biaxial anisotropies and Dzyaloshinskii-Moriya interactions. The spin wave dispersion in an AFM manifests as a bifurcated spectrum with distinct high-frequency and low-frequency bands. Utilizing a heterostructure comprised of platinum and the AFM, we demonstrate anisotropic control of spin-wave bands via spin currents with three-dimensional spin polarizations, encompassing both resonant and propagating wave modes. Moreover, leveraging the confined geometry, we explore the possibility of controlling spin waves within a spectral domain ranging from tens of gigahertz to sub-terahertz frequencies. The implications of our findings suggest the potential for developing a terahertz wave source with electrical tunability, thereby facilitating its incorporation into ultrafast, broadband, and wireless communication technologies.

# INTRODUCTION

The terahertz (THz) gap ranges from 0.3 to 3 THz. This frequency range is of interest in many fields such as the next generation of communication technology [1, 2], and material science [3] and biomedicine [4], security applications [5]. Following decades of research on the gigahertz-frequency dynamics in ferromagnetic materials [6-9], antiferromagnets (AFMs), characterized by two spin sublattices aligned antiparallelly due to negative exchange interactions, are emerging as promising candidates to fill the THz gap in the field of magnetism [10-16]. Recent studies have particularly focused on antiferromagnetic insulators (AFIs), which, due to their strong exchange energy and low damping constant, demonstrate terahertz precession and long magnon propagation length over micrometers [17-21]. In addition to exploiting the inherent properties of AFIs, the ability to control magnetic properties through external stimuli substantially enhances the functionality of AFM-based applications. The capability for tunability of operating frequencies via electrical means is crucial in high-frequency oscillators. Self-oscillation mode triggered by spin-orbit torque (SOT) has been studied theoretically in AFMs [12-16], where the working frequency is linearly modulated as a function of current density. However, entering self-oscillation mode requires higher current densities, which can lead to undesirable thermal effects and should be minimized as much as possible.

In this study, we introduce a broad frequency tuning by manipulating the spin wave band structure via spin-orbit torque (SOT) before the onset of self-oscillation. Our method exploits resonant wave (wavevector, $k = 0$) and propagating wave ($k \neq 0$), enabling the frequency tuning from tens of gigahertz to sub-terahertz frequency. Additionally, we explore the anisotropic control of the spin wave by introducing arbitrary spin polarization directions in three-dimensional space, which has not been explored yet.

Figure 1 shows a schematic where a layered Hall bar structure is composed of a ferromagnet (FM), normal metal (NM) such as polycrystalline platinum (Pt), and a canted AFM $YFeO_3$. Upon applying current, various mechanisms contribute to the generation of a spin current, including the spin Hall effect [8] and anomalous Hall effect [22, 23], spin swapping [24], and spin filtering and precession [25]. Thus, we hypothesize the generation of a spin current with three-dimensional spin polarization because the spin Hall effect is responsible for generating spin currents with in-plane polarization depending on current direction. Other mechanisms are responsible for spin currents with out-of-plane components [22-25]. These accumulated spins impose SOT on the magnetic alignment, leading to an anisotropic modulation in the spin wave dispersion of YFeO$_3$.

Our theoretical analysis reveals a SOT coupling between two distinct spin wave bands: high-frequency (HF) and low-frequency (LF) bands. For example, when the spin current polarization **p** aligns parallel to the Néel vector **l**, the SOT is a spring connecting the two spin wave bands. As a result, frequency is modulated while transferring energy between the LF and HF bands. When **p** is perpendicular to **l**, SOT reduces effective anisotropy or DM interaction energy, reducing spin wave frequency. In a confined geometry, a propagating spin wave gives rise to the formation of a standing wave. The modulated spin waves are excited by transient optical spin current [26], where a circularly polarized femtosecond laser generates the spin current pulse. Therefore, the excitation and detection of sub-THz magnons can be achieved by ultrafast optical pump-probe technique [10, 11, 24] or THz spectroscopy [11], as shown in Fig. 1.

The magnetic system of YFeO$_3$ is of biaxial anisotropies (e.g., the easy axis along the **x** axis and hard axis along the **y** axis) and a Dzyaloshinskii Moriya (DM) interaction where the DM vector is along the **y** axis. The super-exchange interaction occurs between two Fe$^{3+}$ ions, separated by an O$^{2-}$ ion. Below the Néel temperature $T_N$ = 644 K [27], the Fe spins are arranged antiferromagnetically with an easy axis along the **x** direction of the crystal, and DM interaction causes a small canting (~0.4 degrees) of **M**$_1$ and **M**$_2$ along the orthorhombic **c** direction (Fig. 1)[28].

## RESULT AND DISCUSSION

**Equation of motion**

We consider a uniform antiferromagnetic texture composed of YFeO$_3$. To begin, we define total magnetization **m** = (**M**$_1$+**M**$_2$)/2 and the Néel order **l** = (**M**$_1$-**M**$_2$)/2 where **M**$_1$ = **M**$_1$/$M_0$ (**M**$_2$ = **M**$_2$/$M_0$) is normalized by the magnitude of the magnetization $M_0$ = |**M**$_1$| (= |**M**$_2$|). Next, we discuss the energetics of canted antiferromagnet regarding **m** and **l**. The leading-order free energy in the continuum approximation is given as

$$U = \int [\frac{a}{2}|\mathbf{m}|^2 + \frac{A}{2}\frac{\partial^2 \mathbf{l}}{\partial z^2} - \frac{1}{2}K_x l_x^2 - \frac{1}{2}K_z l_z^2 + \frac{\mathbf{D}}{2}\cdot(\mathbf{m}\times\mathbf{l})]d\mathbf{r} \quad (1)$$

where $d\mathbf{l}/dz = (\mathbf{l}_{z+1} - \mathbf{l}_z)/\Delta \sim \mathbf{l}'$ and $\Delta$ is the interspacing of the nearest neighbors along the z axis. Without otherwise stated, thermal fluctuations will be ignored throughout this work. From the left side of Eq. (1), $a$ and $A$ are the homogeneous and inhomogeneous exchange constants[29, 30], $D$ is the homogeneous DM interaction [19], $K_x$ ($K_z$) is the uniaxial anisotropy constant along $\mathbf{x}$ ($\mathbf{z}$) axis.

The functional derivative of energy density is regarded as an effective magnetic field to the lowest order: $\mathbf{H}_{\text{eff,l}} \equiv -\partial U/\partial \mathbf{l} = A(\partial^2 \mathbf{l}/\partial z^2) + K_x l_x \hat{\mathbf{x}} + K_z l_z \hat{\mathbf{z}}$ and $\mathbf{H}_{\text{eff,m}} \equiv -\partial U/\partial \mathbf{m} = -a\mathbf{m} + \mathbf{l} \times \mathbf{D}$, respectively. Regarding the linear terms, the equations of motion are given as [31]

$$\dot{\mathbf{l}} = (\gamma \mathbf{H}_{\text{eff,m}} - \alpha^* \dot{\mathbf{m}}) + \mathbf{T}_{\text{SOT}}^{\text{n}} \quad (2a)$$

$$\dot{\mathbf{m}} = (\gamma \mathbf{H}_{\text{eff,l}} - \alpha \dot{\mathbf{l}}) + \mathbf{T}_{\text{SOT}}^{\text{m}} \quad (2b)$$

where $\dot{\mathbf{l}} = d\mathbf{l}/dt$. $\alpha$ and $\alpha^*$ are damping parameters, and $\gamma$ is a gyromagnetic ratio. Damping-like SOT are given as $\mathbf{T}_{\text{SOT}}^{\text{l}} = \gamma H_{\text{DT}}(\mathbf{m} \times (\mathbf{l} \times \mathbf{p}) + \mathbf{l} \times (\mathbf{m} \times \mathbf{p}))$ and $\mathbf{T}_{\text{SOT}}^{\text{m}} = \gamma H_{\text{DT}}(\mathbf{m} \times (\mathbf{m} \times \mathbf{p}) + \mathbf{l} \times (\mathbf{l} \times \mathbf{p}))$ [32], where $H_{\text{DT}}(=J_c \sigma/\gamma)$ denotes the effective fields corresponding to a damping-like component of SOT torque, $\sigma(=\frac{\gamma \hbar \theta_H}{2etM_s})$ is SOT strength, $J_c$ is charge current density, $M_s$ is the saturation magnetization (Am$^{-1}$), $t$ is the thickness (nm) of the YFeO$_3$, $\theta_H$ is the effective spin Hall angle, $\mathbf{p} = (\sin p_\theta \cos p_\varphi, \sin p_\theta \sin p_\varphi, \cos p_\theta)$ is the DC spin current polarization. The $\mathbf{m}$ can be expressed regarding $\mathbf{l}$ by taking the cross product of $\mathbf{l}$ to equations 2a:

$$\mathbf{m} \sim \frac{1}{a}(\dot{\mathbf{l}} \times \mathbf{l} - \mathbf{D} \times \mathbf{l}) \quad (3)$$

By inserting Eq. 3 into Eq. 2b, we obtain the equation of motion regarding $\mathbf{l}$;

$$\mathbf{l} \times \left(-\ddot{\mathbf{l}} + \nabla \cdot \mathbf{c}^2 \nabla \mathbf{l} + \gamma^2 (K_x l_x \hat{\mathbf{x}} + K_z l_z \hat{\mathbf{z}} + D/a(l_z \dot{\mathbf{l}} - \dot{l}_z \mathbf{l} - l_z \mathbf{D} - D\mathbf{l})) - \beta \dot{\mathbf{l}}\right) = \gamma^2 H_{\text{DT}}(\mathbf{l} \times \mathbf{p}) \times \mathbf{l} \quad , (4)$$

where $\mathbf{c} = (c_a, c_b, c_c)$ is the diagonal tensor of the limiting spin wave velocity along different crystallographic axes.

**Resonant mode**

We consider the $k = 0$ spin wave mode, introducing new collective coordinates $\mathbf{l} = (l_x, l_y, l_z) = (\sin\theta(t)\cos\varphi(t), \sin\theta(t)\sin\varphi(t), \cos\theta(t))$ where $\theta$ and $\varphi$ are polar and azimuthal angles of $\mathbf{l}$. By inserting $\mathbf{l}$ to Eq. (4) and disregarding the nonlinear term, we can obtain the coupled processional mode of $\varphi(t)$ and $\theta(t)$:

$$\ddot{\theta} + 2\omega_E \beta \dot{\theta} + \left(\omega_{HF,0}^2 \sin(\varphi)^2 - \omega_{LF,0}^2\right)\sin(2\theta)/2 - 2\omega_E a\gamma B_{DT}\sin(p_\varphi)\sin(p_\varphi - \varphi) = 0, \quad (5)$$

$$\ddot{\varphi} + 2\omega_E \beta \dot{\varphi} + \omega_{HF,0}^2 \sin(2\varphi)/2 - 2\omega_E a\gamma B_{DT}\left(\cos(p_\theta) - \cos(p_\theta - \varphi)\sin(p_\theta)\cot(\varphi)\right) = 0, \quad (6)$$

where the resonant frequencies of HF and LF modes are defined as $\omega_{HF,0} = 2\omega_E(\omega_{K,x} - \omega_{K,z})$ and $\omega_{LF,0} = 2\omega_E\omega_{K,x} + \omega_D^2$, respectively. In Eqs. 5 and 6, the characteristic angular frequency $\omega(= 2\pi f = \gamma H_{eff})$ is defined as the multiplication of the effective magnetic field $H_{eff}$ and $\gamma$. For example, the exchange field $H_E$, the DMI field $H_D$ and the anisotropy field $H_{K,z}$ ($H_{K,x}$) are defined as $H_E = \omega_E/\gamma$, $H_D = \omega_D/\gamma$, $H_{K,x} = \omega_{K,x}/\gamma$ ($H_{K,z} = \omega_{K,z}/\gamma$).

**Spin wave dispersion**

Under applying DC spin current, we confirm that the $\mathbf{l}$ is uniformly aligned. Therefore, we can choose spin wave ansatz, $\mathbf{l} = (\mathbf{l}_r, \mathbf{l}_\varphi A_\varphi e^{i(\omega t - kx)}, \mathbf{l}_\theta A_\theta e^{i(\omega t - kx)})$ where $\mathbf{l}_r = (\sin\theta_0\cos\varphi_0, \sin\theta_0\sin\varphi_0, \cos\theta_0)$, $\mathbf{l}_\varphi = (-\sin\varphi_0, \cos\varphi_0, 0)$, $\mathbf{l}_\theta = (\cos\theta_0\cos\varphi_0, \cos\theta_0\sin\varphi_0, -\sin\theta_0)$. $A_\varphi$ and $A_\theta$ are spin wave amplitude, $\varphi_0$ and $\theta_0$ are equilibrium angles of $\mathbf{l}$ under the SOT. Then, we can obtain a dispersion relation $\omega^2 = c_c^2 k^2 + \omega_0^2$, where $\omega_0$ is the resonant frequency such as $\omega_{0,HF}$ and $\omega_{0,LF}$, $k(= 2\pi q)$ is the spin wave vector and $c$ is limiting group velocity. Note that the spin wave dispersion and propagation velocity, $\upsilon(J_c) = \frac{\partial \omega}{\partial k}$ are affected by electrically-modulated resonant frequency ($\omega(J_c) = \sqrt{c^2 k^2 + \omega_0(J_c)}$). In numerical and analytical calculation, we choose the following parameters for YFeO$_3$ where $\omega_E/(2\pi) = 18.2$ THz ($H_E = 650$ T), $\omega_D/(2\pi) = 336$ GHz ($H_D = 12$ T), $\omega_{Kx}/(2\pi) = 4.5$ GHz ($H_{Kx} = 160$ mT), and $\omega_{Kz}/(2\pi) = 2.0$ GHz ($H_{Kz} = 71$ mT) [33, 34] and $c_c = 38$ km s$^{-1}$ along $\mathbf{c}$ axis [35]. The effective damping-like SOT field $H_{DT} = J_c \frac{\hbar\theta_H}{2etM_s}$ is given by the reduced Planck constant $\hbar = 1.054 \times 10^{-34}$ J s, the spin Hall angle $\theta_H = 0.08$, the electron charge $e$, the effective penetration depth $t$ of spin current into YFeO$_3$ of 4 nm, $M_s$ of 200 A m$^{-1}$ and $\beta = 5\times10^{-4}$ [28]. The femtosecond laser is incident at an angle $\delta = 45$ degrees. Consequently, the optical spin current introduced has polarizations set at 45-degree angles between the **y** and **z** axes, activating both HF and LF modes. In this work, the field-like torque of spin current is not considered to show dynamics clearly, and the amplitude of a generated spin wave is very weak ($\varphi(t) \ll 1$, $\theta(t) \ll 1$). The resonance frequencies for LF and HF modes are obtained as $\omega_{\text{LF},0}/(2\pi) = \sqrt{(2\omega_E\omega_{K,x} + \omega_D^2)}/(2\pi) = 300$ GHz and $\omega_{\text{HF},0}/(2\pi) = \sqrt{2\omega_E(\omega_{K,x} - \omega_{K,z})}/(2\pi) = 525$ GHz, respectively. To achieve enhanced resolution in dispersion relation, we employed a linearized dispersion relation incorporating mesh and adjusted magnetic parameters in the numerical calculations (see Methods and Supplementary Notes 1).

**Current density dependence for three extreme cases: ($p_\theta$, $p_\varphi$) = (0, $\pi$/2) or ($\pi$/2, $\pi$/2) or (0, 0)**

As we consider biaxial AFMs (i.e., the two easy axes in the **x** and **z** direction), ($p_\theta$, $p_\varphi$) = ($\pi$/2, 0), ($\pi$/2, $\pi$/2) and (0,0) correspond to the case where **p**//**x**, **p**//**y**, and **p**//**z**, respectively. In the subsequent analysis, we applied $J_c$ of 2.5, 1.2, and 3.5 A m$^{-2}$ for **p**//**x**, **p**//**y**, and **p**//**z** orientation, respectively, which are chosen as the current densities just before self-oscillation (see insets of Figs. 2b, 2d and 2f). This differentiation in $J_c$ values is considered to highlight the variations in frequency distinctly.

First, we consider the case for ($p_\theta$, $p_\varphi$) = ($\pi$/2, 0) (i.e., **p**//**x**). Assuming weak excitation ($\theta(t) \ll 1$ and $\varphi(t) \ll 1$), Eqs. 5 and 6 are linearized as

$$\ddot{\theta} + 2\omega_E\beta\dot{\theta} + (\omega_{\text{LF}})^2\theta + \Omega\varphi = 0, \quad (7)$$

$$\ddot{\varphi} + 2\omega_E\beta\dot{\varphi} + (\omega_{\text{HF}})^2\varphi - \Omega\theta = 0, \quad (8)$$

where $\Omega = 2\omega_E\gamma H_{\text{DT}}$ is the coupling constant connecting HF and LF bands. Linearized Eqs. 5 and 6 are reminiscent of the two pendula systems coupled by spring (see Fig. 2a), where some of the energy that is contained within the first pendulum is transferred to the second one and vice versa through the SOT spring. Here, the transfer power $\Omega = 2\omega_E\gamma H_{\text{DT}}$ is proportional to the spin current density. Two modes are decoupled for $J_c = 0$ A m$^{-2}$. They are selectively driven depending on the driving force polarizations (HF mode for **p**//**z** and LF mode for **p**//**y** [34].

However, as spin current is applied (see Fig. 2b), HF (LF) band frequency decreases (increase). By solving Eqs. 7 and 8, the dispersion relations are given as

$$\omega_{LF,k} = -i\omega_E\beta + \frac{1}{2}\sqrt{(2c_ck)^2 + (\omega_{LF,0}^2 + \omega_{HF,0}^2) - \sqrt{(\omega_{HF,0}^2 - \omega_{LF,0}^2)^2 - (4\omega_E\gamma H_{DT})^2}}$$

and

$$\omega_{HF,k} = -i\omega_E\beta + \frac{1}{2}\sqrt{(2c_ck)^2 + (\omega_{LF,0}^2 + \omega_{HF,0}^2) + \sqrt{(\omega_{HF,0}^2 - \omega_{LF,0}^2)^2 - (4\omega_E\gamma H_{DT})^2}}$$. As shown in Fig.

2b, as wavevector $k$ is large, frequency modulation is less efficient, implying that DC current efficiently modulates spin waves around $q \sim 0$. For example, for $J_c = 2.5 \times 10^{12}$ A m$^{-2}$ or $H_{DT}(=J_c\sigma/\gamma) = 0.52$ T, the frequency modulation $\Delta f = (f - f_0)$ of LF (HF) mode is +85.2 GHz (−58.4 GHz) at $q = 0$ nm$^{-1}$ while $\Delta f$ of LF (HF) mode frequency is +40 GHz (−36 GHz) at $q = 0.0167$ nm$^{-1}$. Note that two modes become overlapped at critical current density $J_{c,self} = \frac{\omega_D^2 + 2\omega_{Kz}\omega_E}{4\omega_E\sigma} \sim 2.73 \times 10^{12}$ A m$^{-2}$ or $H_{DT,self} = 0.56$ T (see the inset of Fig. 2b). Here, $J_{c,self}$ is obtained in the condition that intrinsic damping and SOT are canceled out. Alternatively, it is given by setting an imaginary part of $\omega_{LF,k}^2$ and $\omega_{HF,k}^2$ equal to zero. At $J_c > J_{c,self}$, all spin waves become damping-less and go into self-oscillation mode.

Second, we consider the case for $(p_\theta, p_\varphi) = (\pi/2, \pi/2)$ (i.e., **p**//**y**). Equations of motions are simplified as

$$\ddot{\theta} + 2\beta\omega_E\dot{\theta} + (\omega_{0,LF}^2 - \omega_{0,HF}^2\sin(\varphi)^2)\sin(2\theta)/2 - 2\omega_E\gamma H_{DT}\cos(\varphi) = 0 \quad (9)$$

$$\ddot{\varphi} + 2\beta\omega_E\dot{\varphi} + \sin(2\varphi)\omega_{0,HF}^2/2 - 2\omega_E\gamma H_{DT}\sin(\varphi)\cot(\theta) = 0 \quad (10)$$

As sublattice dynamics (i.e., **m**$_1$(t) and **m**$_2$(t)) are considered, the $J_c$ induces temporally spin canting of **m**$_1$ and **m**$_2$ along the **y** axis. Subsequently, the enhanced exchange energy partially transfers it to anisotropy, reorienting the equilibrium **l** into $\theta \rightarrow \theta + \theta_0$. As a result, **l** is reoriented on the **xz** plane (see the green arrow in Fig. 2c). Therefore, applying $J_c$ changes the equilibrium angles as $\theta_0 = (\pi - \arcsin(4\gamma H_{DT}/\omega_{LF}^2))/2$ and $\varphi_0 = 0$. By rotating the reference frame $\theta \rightarrow \theta + \theta_0$, Eq. 9 is recast as $\ddot{\theta} + 2\beta\omega_E\dot{\theta} + \omega_{0,LF}^2\cos(2\theta_0)\sin(2\theta)/2 - 2\omega_E\gamma H_{DT} = 0$. Therefore, the resonant frequencies of LF mode are modulated efficiently by the following relation: $\omega_{LF} = \omega_{0,LF}\sqrt{|\cos(2\theta_0)|}$ and accordingly, $\omega_{HF} = \sqrt{\omega_{0,HF}^2 - 2\gamma H_{DT}\omega_E\cot\theta_0}$. Therefore, the spin wave dispersion is given by $\omega_{HF,k} = \sqrt{c_c^2k^2 + \omega_{HF}^2}$ and $\omega_{LF,k} = \sqrt{c_c^2k^2 + \omega_{LF}^2}$. For example, for $J_c$

= 1.2 × 10$^{12}$ A m$^{-2}$ or $H_{DT,self}$ = 0.25 T, the frequency modulation of LF (HF) mode is -106 GHz (-25.8 GHz) at $q$ = 0 nm$^{-1}$ while $\Delta f$ of LF (HF) mode frequency is -38.64 GHz (-16 GHz) at $q$ = 0.0167 nm$^{-1}$ (see Fig. 3d).

Third, we consider the case for $(p_\theta, p_\varphi)$ = (0, 0) (i.e., **p**//**z**). Equations of motions are given as

$$\ddot{\theta} + 2\beta\omega_E\dot{\theta} - \sin(2\theta)(\omega_{HF}^2 \sin(\varphi)^2 - \omega_{LF}^2)/2 = 0 \tag{11}$$

$$\ddot{\varphi} + 2\beta\omega_E\dot{\varphi} + \sin(2\varphi)\omega_{HF}^2/2 - 2\omega_E\gamma H_{DT} = 0 \tag{12}$$

The SOT induces temporally spin canting of **m**$_1$ and **m**$_2$ along the **z** axis, increasing the exchange energy. Subsequently, the enhanced exchange energy partially transfers it to anisotropy, reorienting the equilibrium **l** vector into $\varphi \to \varphi+\varphi_0$ where the reorientation angle is defined as $\varphi_0 = \left(\pi - \arcsin(4\gamma H_{DT} / \omega_{HF}^2)\right)/2$ (see the green arrow in Fig. 2b). In the new reference frame, Eq. 12 is recast as $\ddot{\varphi} + 2\beta\omega_E\dot{\varphi} + \omega_{0,HF}^2 \cos(2\varphi_0)\sin(2\varphi)/2 + 2\gamma H_{DT}(\cos(2\varphi) - \omega_E) = 0$. One can see that the resonant frequencies of HF mode are modulated efficiently by the following relation: $\omega_{HF} = \omega_{0,HF}\sqrt{|\cos(2\varphi_0)|}$ and accordingly, $\omega_{LF} = \sqrt{\omega_{0,HF}^2 \sin(\varphi_0)^2 - \omega_{0,LF}^2}$. Accordingly, spin wave dispersion is obtained by $\omega_{HF,k} = \sqrt{c_c^2 k^2 + \omega_{HF}^2}$ and $\omega_{LF,k} = \sqrt{c_c^2 k^2 + \omega_{LF}^2}$, as shown in Fig. 3e. For example, for $J_c$ = 3.5 × 10$^{12}$ A m$^{-2}$ or $H_{DT,self}$ = 0.72 T, the frequency modulation of LF (HF) mode $\Delta f$ is -158.1 GHz (-157.3 GHz) at $q$ = 0 nm$^{-1}$ while $\Delta f$ of LF (HF) mode frequency is -52.1 GHz (-90 GHz) at $q$ = 0.0167 nm$^{-1}$ (see Fig. 3f). Note that at $\varphi = \varphi_{0,SW}$, the critical current density for switching is defined as $J_{SW} = \omega_{HF}^2 \sin\left(2\arcsin\left(\omega_{LF}/\omega_{HF}\right)\right)/4\omega_E\sigma\hbar$. When $\varphi > \varphi_{0,SW}$, the $\theta_0$ is switched from $\pi/2$ to 0, and the frequency increases with relation: $\omega_{LF} = \sqrt{-\omega_{0,HF}^2 \sin(\varphi_0)^2 + \omega_{0,LF}^2}$ (see insets of Figs. 2e and 2f).

Why this change is different.

**Arbitrary polarization directions: $(p_\theta, p_\varphi)$ = ($0 < p_\theta < \pi/2$, $0 < p_\varphi < \pi/2$).**

Figure 3 presents the polarization dependence for modes at $q$ values of 0 and 0.0167 nm$^{-1}$ under $J_c$ of 1.2×10$^{12}$ A m$^{-2}$. This current density was selected as it represents the lowest current that induces self-oscillation among the cases for **p**//**x**, **p**//**y** and **p**//**z** (see the inset of Fig. 2d). We define tunability as $\varepsilon(p_\theta, p_\varphi) = \Delta f(p_\theta, p_\varphi)/f_0 \times 100$. The $\varepsilon(p_\theta, p_\varphi)$ explains two cases. For **p**//**y** and **p**//**z** cases, the variation of $\varphi$ and $\theta$ by the spin current corresponds to $\Delta f$. In other words, it is similar to how easily the oscillator overcomes the potential barrier $V_B$, which is defined as $V_{B,LF} \equiv -\int \omega_{LF}^2 \sin(2\theta)/2 d\theta$ and $V_{B,HF} \equiv -\int \omega_{HF}^2 \sin(2\varphi)/2 d\varphi$ (i.e., $|V_{B,LF}| < |V_{B,HF}|$ for YFeO$_3$). For **p**//**x**, how efficiently spin current transfers energy between LF and HF modes determines $\Delta f$. As a result, $|\varepsilon(\pi/2, 0)| > |\varepsilon(\pi/2, \pi/2)| > |\varepsilon(0, 0)|$ (i.e., $|\varepsilon|$ at **p**//**y** > $|\varepsilon|$ at **p**//**x** > $|\varepsilon|$ at **p**//**z**); the $\varepsilon$ ranges from $\varepsilon(\pi/2, 0) = -34$ % to $\varepsilon(0, 0) = +5.6$ % for LF mode and from $\varepsilon(\pi/2, 0) = -4.8$ % to $\varepsilon(\pi/2, 0) = -1.8$ % for HF mode at $q = 0$, while from $\varepsilon(\pi/2, 0) = -5.4$ % to $\varepsilon(0, 0) = 1$ % for LF mode and from $\varepsilon(\pi/2, 0) = -2.0$ % to $\varepsilon(0, 0) = -0.8$ % for HF mode at $q = 0.0167$ nm$^{-1}$.

**Total magnetization dynamics: m(t)**

As written by Eq. (3), the Néel order dynamics reproduce the total magnetization dynamics. However, the nonlinear coupling between **m** and **l** is not calculated analytically. Therefore, we performed the numerical simulation where the alternating signs of DM interaction in space are necessary to form the uniformed canted antiferromagnet (see Supplementary Note 1). Next, we performed the numerical calculation using techniques of mesh-grid with mesh number $N$ and energetic adjustment to reduce the simulation time where $N = 10$, $t = 30$ nm, and $q = 0.0167$ nm$^{-1}$, as shown in Fig. 4 (see Supplementary Note 2). The optically injected spin current with the polarization of (0,1,1), oriented orthogonally to **l**, exerts a torque on the **l** vector (1,0,0) (see Fig. 1). As a result of the impulsive optical SOT, the tilted **m** starts precessing, leading to circularly polarized electromagnetic emission $\mathbf{E}_{THz} \propto \partial^2 \mathbf{m}/\partial t^2$ from the rotating magnetic dipole [11, 28, 34].

Without applying spin current, the **m**'s precession in LF (HF) mode is defined by Eq. 3, and trajectories of **m** are described by precessional (oscillating) motion around (on the **xy** plane) the **z** axis [11, 28, 34] (see Figs. 4a-1 and 4a-2). For example, for $J_c = 0$, $m_x$ (black line) and $m_y$ (red line) show $q = 0$ LF mode at 300 GHz and standing wave mode at 703 GHz, while $m_z$ (blue line) shows $q = 0$ HF mode at 525 GHz and standing wave mode at 824 GHz (see Fig. 4a).

At **p**//**x** and $J_c = 2.5 \times 10^{12}$ A m$^{-2}$, the spectrum of **m**'s precession exhibits coupled modes and frequency change of **m**'s precession is coincident with that calculated in the Néel order dynamics as shown in Fig. 2b (see Fig. 4b). At **p**//**y** and $J_c = 1.2 \times 10^{12}$ A m$^{-2}$, the spectra of $m_x$ and $m_z$ show unexpectedly coupled modes different from $m_y$. In our analysis, the coupled modes of $m_x$ and $m_z$ arise due to an additional SOT effect on **m** where in **p**⊥**m**, high-order terms, not accounted for in our analytical method, should be considered; in collinear AFM without DMI, $m_x$, $m_y$ and $m_z$ show decoupled modes for **p**//**y** (see Supplementary Note 3). At **p**//**z** and $J_c = 3.5 \times 10^{12}$ A m$^{-2}$, the spectrum of **m**'s precession shows decoupled modes where DC SOT does not exert additional torque to **m** because **p**//**m**.

Finally, it is worth comparing the frequency modulation efficiency between SOT and external magnetic field $H_{ext}$. We observed that frequency tuning by spin current exhibits a nonlinear behavior. This is in contrast to the Zeeman field-frequency modulation, which demonstrates a linear tuning pattern [36, 37]. Notably, higher current densities yield significantly greater frequency tuning efficiency than magnetic fields. For example, at **p**//**y**, $J_c = 2.5 \times 10^{12}$ A m$^{-2}$ is converted to magnetic field $H_{DT} = 0.52$ T, leading to $\Delta f = 85.2$ GHz ($k = 0$ LF mode) and 58.4 GHz ($q = 0$ HF mode). This $\Delta f$ is four times more efficient than 14.84 GHz by $H_{ext} = 0.52$ T, estimated from $\gamma$ (= 28 GHz T$^{-1}$). For any polarization (**p**//**z** or **p**//**y**), $\Delta f$ is more efficient than that by an external magnetic field. One key to realizing the spin wave modulation by a spin current with arbitrary spin polarizations is to pattern the Hall bar (nanometer scale). This could be performed by sophisticated deposition or lithography techniques on the atomic scales.

## IV. CONCLUSION

This work investigated spin wave manipulation by the spin current in a biaxial antiferromagnet. The damping-like SOT is found to be an efficient way to manipulate spin wave frequency. Investigations reveal that the dynamics influenced by SOT are analogized to two pendulums coupled distinctly by SOT, contingent upon spin polarization direction; frequency modulation efficiency depends on the height of the potential barrier for **p** ⊥ **l** or energy transfer power of SOT for **p** ∥ **l**. The modulated frequency ranges from tens of gigahertz to the sub-terahertz frequency range, including the standing wave. Our method is more efficient than an external magnetic field in frequency tuning. The coupling between **m** and **l** leads to THz electromagnetic radiation by **m**'s precession, controlled by spin current.

**METHODS**

**Numerical simulation**

Numerical simulation using Mathematica is made to solve the Landau-Lifshitz-Gilbert equation for the micromagnetic calculations.

$$\frac{d\mathbf{M}_i}{dt} = -\gamma\left(\mathbf{M}_i \times \mathbf{H}_{eff} + \beta \mathbf{M}_i \times (\mathbf{M}_i \times \mathbf{H}_{eff})\right) + \mathbf{M}_i \times (\mathbf{M}_i \times \mathbf{\Gamma}_i),$$

for every simulation cell $\mathbf{M}_i$ of the discretized vector field of the magnetization. Here, $\gamma = 1.76 \times 10^{11}$ T$^{-1}$ s$^{-1}$ denotes the gyromagnetic ratio of an electron, $\beta$ is the Gilbert damping parameter. The time- and space-dependent effective magnetic field,

$$\mathbf{H}_{eff}^i(t,z) = \mathbf{H}_E^i + \mathbf{H}_D^i + \mathbf{H}_{Kx}^i + \mathbf{H}_{Kz}^i + \mathbf{H}_{ext}^i,$$

is composed of the external field, $\mathbf{H}_{ext}$; exchange field $\mathbf{H}_E^i$, a Dzyaloshinskii-Moriya field $\mathbf{H}_D^i$, uniaxial anisotropy field along $\mathbf{x}$ ($\mathbf{z}$) axis $\mathbf{H}_{Kx}^i$ ($\mathbf{H}_{Kz}^i$) and an external magnetic field, $\mathbf{H}_{ext}^i$. In addition, we applied optical spin-orbit torque (SOT) $\Gamma_i$ to $\mathbf{M}_i$ where $\Gamma_i = J_c \sigma \mathbf{p}_i$ is composed of current density, $J_c$, SOT strength $\sigma = \frac{\gamma \hbar \theta_H}{2etM_s}$ (s$^{-1}$), the saturation magnetization $M_s$ (A m$^{-1}$), the thickness $t$ (nm) of the YFeO$_3$, the effective spin Hall angle $\theta_H$ and polarization $\mathbf{p}_i$. To excite LF and HF band by optical spin-orbit torque (OSOT), we applied time and space-dependent SOT, $\Gamma_{ext}(t, z) = 10000$ s$^{-1}$ × sinc($2\pi f_0 t$) × sinc($2\pi q_0 z$) ($\mathbf{y}+\mathbf{z}$) where $f = 40$ THz and $q = 1$ nm$^{-1}$. Numerical simulations were conducted from 0 to 1 μs with a minimum time interval of 0.1 ps.

**Data availability**

All other data that support the plots within this paper and other findings of this study are available from the Supplementary Note or the corresponding author upon reasonable request.

**Acknowledgments**

This research was supported by the National Research Foundation of Korea (NRF) grant funded by the Korea government (MEST) (No. RS-2024-004113774) and by the Nation Research Foundation (NRF) of Korea (Grants No. 2022R1A2C1006504).

**Author contributions**

G.-M.C. and T.H.K. conceived the project idea and planned the analytical and numerical calculations. T.H.K. performed the analytical and numerical calculations. G.-M.C. and T.H.K. analyzed the data. G.-M.C. and T.H.K. led the work and wrote the manuscript. The results of the theoretical and numerical findings were discussed by T.H.K., J.-I.K, G.-J.K, K.-H.K and G.-M.C..

**Competing interests**

The authors declare no competing interests.

**Additional information**

**Extended data** is available for this paper.

**Supplementary information** the online version contains supplementary material available for this paper.

**Correspondence and requests for materials** should be addressed to T.H. Kim and G.-M. Choi.

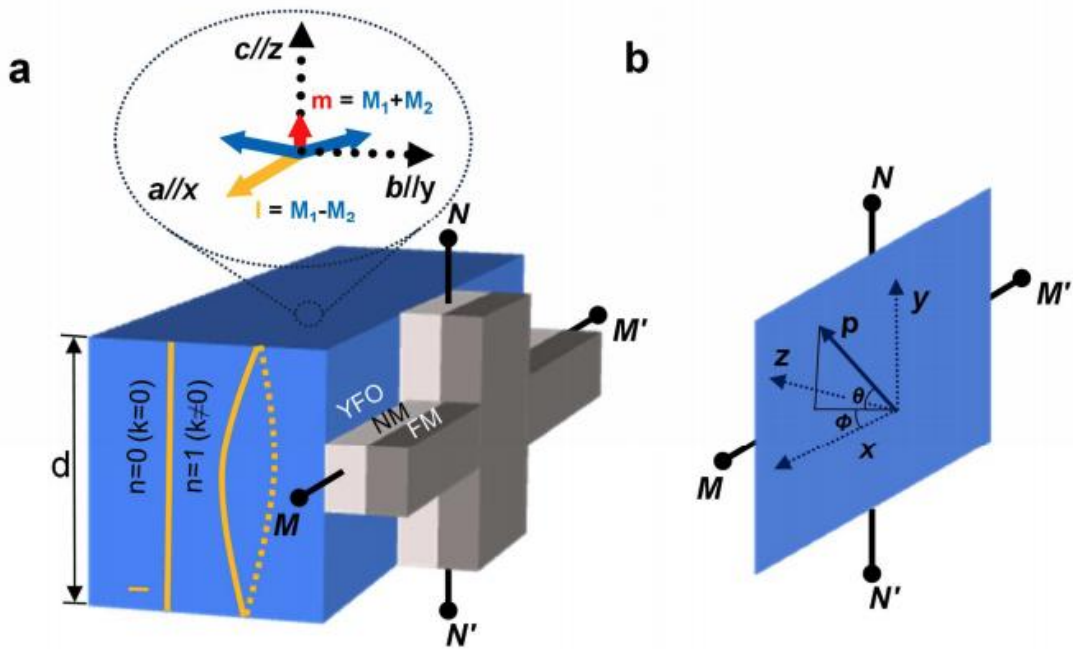

**Fig. 1** | Schematics for electrically modulated spin-wave of YFeO$_3$ where a layered Hall bar structure is composed of a ferromagnet (FM), normal metal (NM) on a canted AFM YFeO$_3$. NM is polycrystalline platinum. (a) DC spin current with arbitrary spin polarization **p** is injected into YFeO$_3$ layer and modulates the spin wave bands of YFeO$_3$. A circularly polarized femtosecond pulse generates an optical spin-orbit current exerting torque on YFeO$_3$, thereby generating a spin wave. Standing and resonant waves generate electromagnetic waves in confined geometry through rotating magnetization, **m**. Inset shows equilibrium state of canted antiferromagnet. (b) The injection of spin curdrent with arbitrary spin polarization directions to YFeO$_3$.

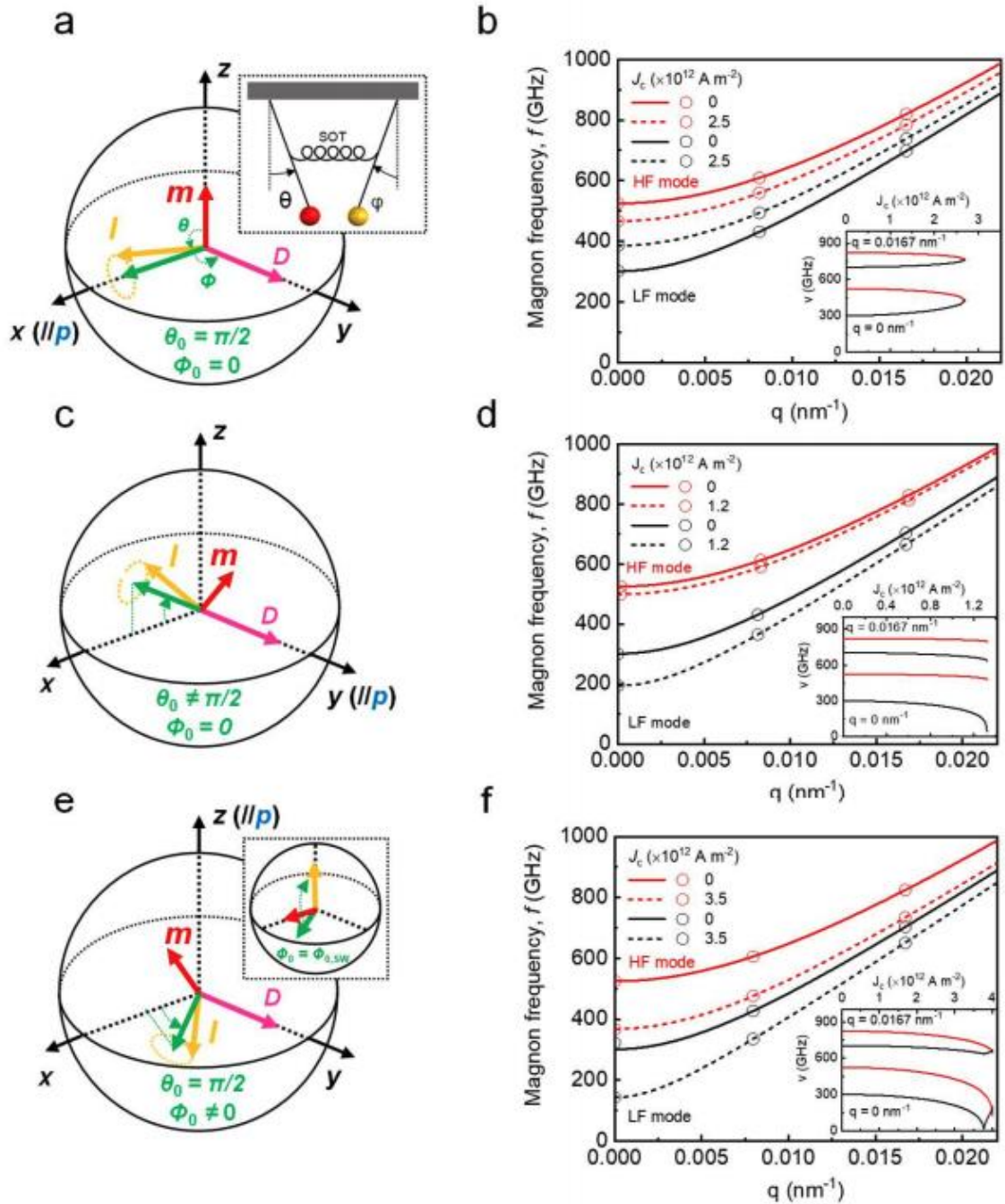

**Fig. 2 |** Spin wave manipulation under applying DC spin current with different polarizations. **a** A schematic for a precessional motion for **l**//**p**//**x**. As shown in the inset, DC SOT plays the role of spring connecting $\theta(t)$ and $\varphi(t)$. **b** Spin wave dispersion relation for **p**//**x**. LF band frequency increases while HF band frequency decreases. The inset shows spin current dependence of HF and LF mode at $q = 0$ and $q = 0.0167$ nm$^{-1}$. **c** A schematic for a precessional motion for **l**⊥**p**(//**y**). DC SOT reorients **l** on the **xz** plane where $\mathbf{l_r} = (\sin\theta_0, 0, \cos\theta_0)$. **d** Spin

wave dispersion relation for **p**//**y**. Both LF and HF band frequencies decrease. The inset shows spin current dependence of HF and LF mode at $q = 0$ and $q = 0.0167$ nm$^{-1}$. **e** A schematic for a precessional motion for **l**⊥**p**(//**z**). DC SOT reorients **l** on the **xz** plane where $\mathbf{l}_r = (\cos\varphi_0, \sin\varphi_0, 0)$. As shown in the inset, $\varphi_{0,SW}$, $l_r$ is switched along the **z** axis at a switching angle. **f** Spin wave dispersion relation for **p**//**z**. Both LF and HF band frequencies decrease. The inset shows spin current dependence of HF and LF mode at $q = 0$ and $q = 0.0167$ nm$^{-1}$. After switching, the LF band frequency increases. In **b**, **d** and **f**, lines and symbols indicate analytical and numerical calculated spin wave dispersion, respectively.

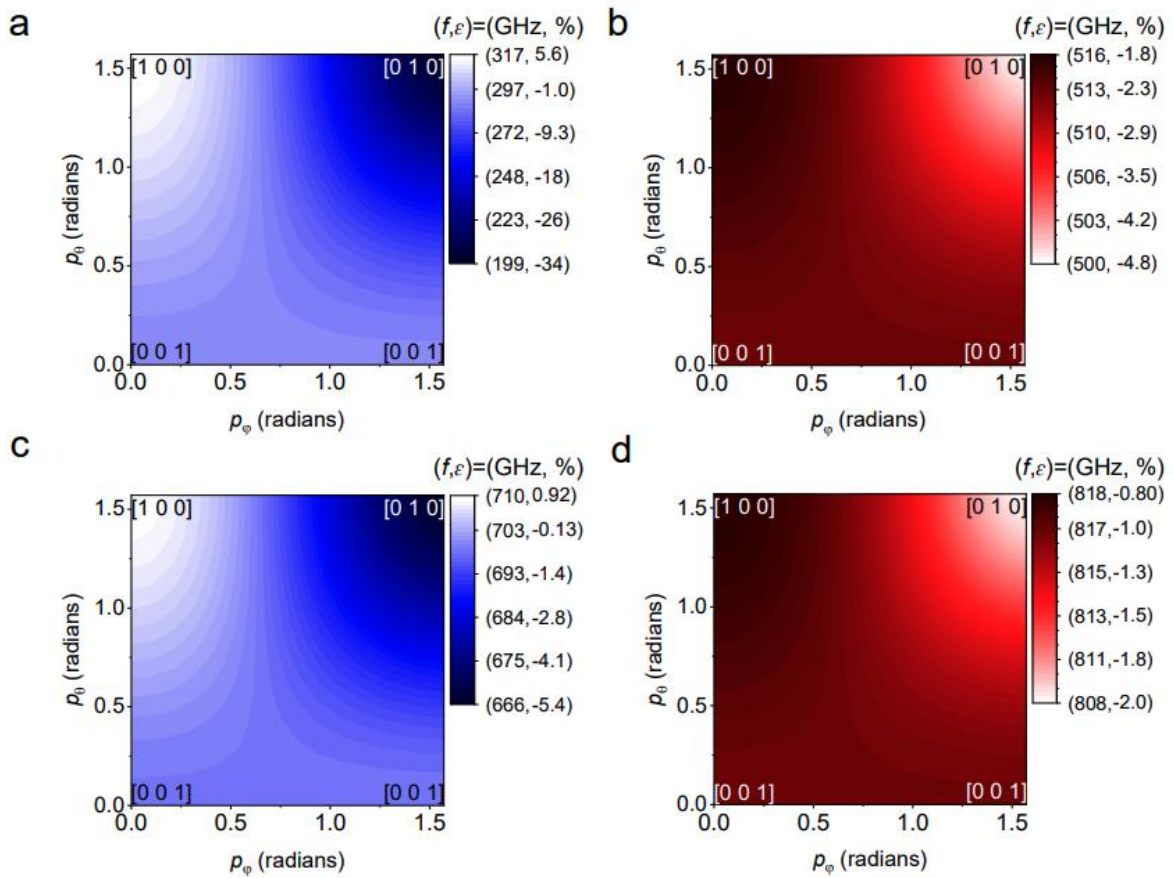

**Fig. 3** | The modulated spin wave frequency depending on different spin current polarizations under applying $J_c = 1.2 \times 10^{12}$ A m$^{-2}$. **a** LF mode at $q = 0$ nm$^{-1}$. **b** HF mode at $q = 0$ nm$^{-1}$. **c** LF mode $q = 0.0167$ nm$^{-1}$. **d** HF mode $q = 0.0167$ nm$^{-1}$. $f$ is modulated frequency and $\varepsilon = \Delta f/f_0 \times 100$

is modulation efficiency. Here, $f_0$ at $q = 0$ is 300 GHz (LF mode) and 525 GHz (HF mode), and $f_0$ at $q = 0.0167$ nm$^{-1}$ is 703.5 GHz (LF mode) and 824.6 GHz (HF mode).

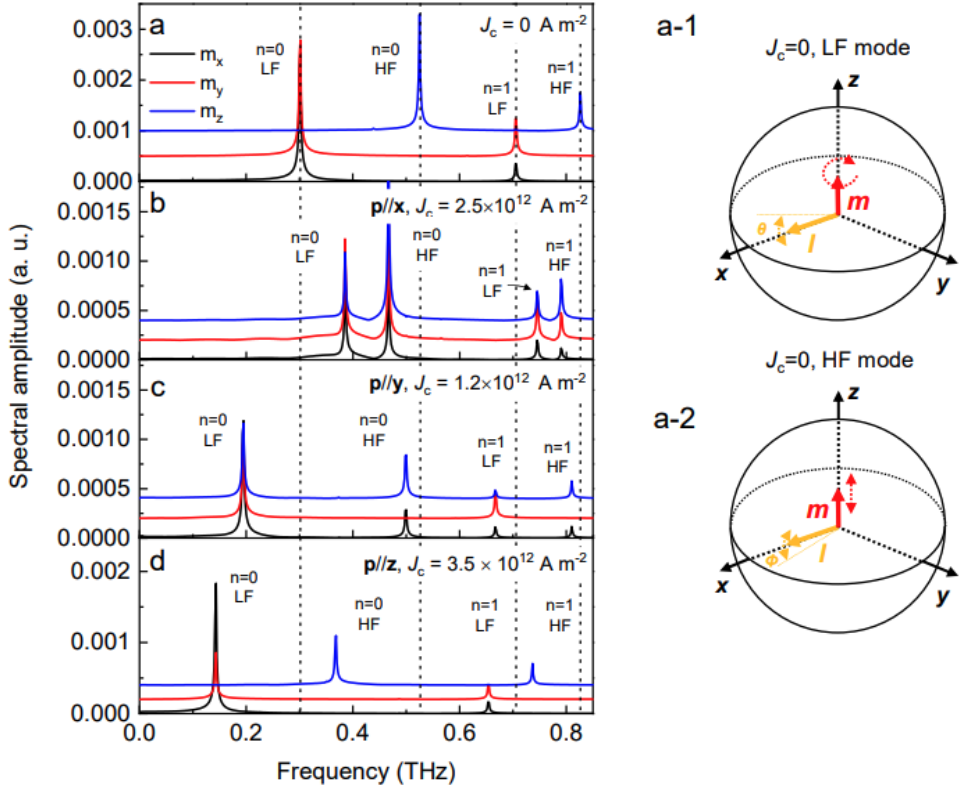

**Fig. 4** | THz emission spectra estimated from **m**'s precession. **a** Spectra without applying spin current. The dynamics of **m** and **l** are described in **a-1** and **a-2**, respectively. **b** Spectra with $J_c = 2.5 \times 10^{12}$ A m$^{-2}$ with **p**//**y**. **c** Spectra with $J_c = 1.2 \times 10^{12}$ A m$^{-2}$ with **p**//**x**. **d** Spectra with a $J_c = 3.5 \times 10^{12}$ A m$^{-2}$ with **p**//**z**. In Figure, $m_y$ and $m_z$ are shifted by 0.0002 and 0.0004. The dotted line indicates LF and HF frequencies at $q = 0$ and $q = 0.0167$ nm$^{-1}$ without applying current; $f_0$ at $q = 0$ is 300 GHz (LF mode) and 525 GHz (HF mode), and $f_0$ at $q = 0.0167$ nm$^{-1}$ is 703.5 GHz (LF mode) and 824.6 GHz (HF mode).